\documentclass{appolb}
\usepackage{graphicx}
\usepackage{latexsym,amssymb,euscript}
\usepackage{xcolor}
\usepackage{amsmath}
\usepackage{cite}
\begin{document}
\title{\bf Leptoproduction of $\rho$-mesons as discriminator for the unintegrated gluon distribution in the proton
\thanks{Presented by A.D. Bolognino at \textbf{\emph {Diffraction and Low-x 2018}}}%
}
\author{\underline{A.D. Bolognino}$^{1,2}$, F.G. Celiberto$^{3,4}$, D.Yu. Ivanov$^{5,6}$, A. Papa$^{1,2}$
\address{\centerline{${}^1$ {\sl Dipartimento di Fisica, Universit{\`a} della Calabria, I-87036 Arcavacata di Rende, Cosenza, Italy}}
	\centerline{${}^2$ {\sl Istituto Nazionale di Fisica Nucleare, Gruppo collegato di Cosenza,}} \centerline{\sl I-87036 Arcavacata di Rende, Cosenza, Italy}
	\centerline{${}^3$ {\sl Dipartimento di Fisica, Universit{\`a} degli Studi di Pavia, I-27100 Pavia, Italy}}
	\centerline{${}^4$ {\sl Istituto Nazionale di Fisica Nucleare, Sezione di Pavia, I-27100 Pavia, Italy}}
	\centerline{${}^5$ {\sl Sobolev Institute of Mathematics, 630090 Novosibirsk, Russia}}
	\centerline{${}^6$ {\sl Novosibirsk State University, 630090 Novosibirsk, Russia}}
}
}
%{Third Author of different affiliation
%the Name(s) of other Author(s)
%\address{affiliation}
\maketitle
\begin{abstract}
\vspace{-0.2cm}
The gluon content of the proton, in the high-energy regime, is embodied by the unintegrated gluon distribution (UGD), which describes the gluon emission probability, with a given longitudinal momentum fraction and transverse momentum. The UGD, formulated within the $\kappa$-factorization approach, has universal validity and several models for it have been proposed so far. We will show that  the polarized $\rho$-meson leptoproduction at HERA is a not trivial testfield for discriminating among existing models of UGD.
\end{abstract}
\vspace{-0.2cm}
\section{Introduction}
An increasingly detailed understanding on the structure of the proton is the fundamental point of our ability to investigate on the dynamics of strong interactions at LHC and find new Physics.
In the description of the collision processes, the information about the inner structure of the proton is enclosed in the partonic distribution functions which enter the factorized expression for the cross section. In the deep inelastic scattering (DIS), featuring high photon virtuality, $Q^2$, and large center-of-mass-energy of the virtual photon-proton system $W$, $W\gg Q\gg \Lambda_{\rm QCD}$, which implies small $x=Q^2/W^2$, the suitable factorization approach is provided by $\kappa$-\emph{factorization}.
Here the DIS cross section becomes
%In the high-energy regime, $W\gg Q\gg \Lambda_{\rm QCD}$, which implies at small $x=Q^2/W^2$, the suitable factorization approach is provided by $\kappa$-\emph{factorization}, where the DIS cross section is defined 
the convolution of the unintegrated gluon distribution (UGD) in the proton with the impact factor (IF) for the $\gamma^* \rightarrow \gamma^*$ transition. The UGD, in its original definition, obeys the BFKL~\cite{BFKL} evolution equation in the $x$ variable and, being a nonperturbative quantity, it is not well known and a good number of models for it have been introduced so far (see, for instance, \cite{small_x_WG}).
The goal is to show that HERA data on polarization observables in vector meson (VM) leptoproduction can be employed to constrain the $\kappa$ dependence of the UGD in the HERA energy domain. The observable under investigation is the ratio of the two dominant amplitudes for the polarized leptoproduction of $\rho$ mesons, namely the longitudinal VM production from longitudinally polarized virtual photons and the transverse VM production from transversely polarized virtual photons. First we illustrate the expression of the dominant amplitudes just mentioned above; then we provide a pattern with the essential details of a few models for UGD and compare theoretical predictions \cite{Bolognino:2018, Bolognino:2018mlw} from the tested models of UGD with HERA data.
\vspace{-0.3cm}
\section{Theoretical framework}
Fruitful and exhaustive analyses of the hard exclusive production of the $\rho$ meson in $ep$ collisions, given by $\gamma^*(\lambda_\gamma)p\rightarrow \rho (\lambda_\rho)p$,
%\begin{equation}
%\label{process}
%\gamma^*(\lambda_\gamma)p\rightarrow \rho (\lambda_\rho)p\,
%\end{equation}
are provided by H1 and ZEUS collaboration, where $\lambda_{\rho,\gamma}$ represent the meson and photon helicities, respectively, and can take the values 0 for the longitudinal polarization and $\pm 1$ transverse ones. The helicity amplitudes 
$T_{\lambda_\rho \lambda_\gamma}$ measured at HERA~\cite{Aaron:2009xp} reveal a peculiar ordering: $T_{00} \gg T_{11} \gg T_{10} \gg T_{01} \gg T_{-11}.$
%\begin{equation}
%T_{00} \gg T_{11} \gg T_{10} \gg T_{01} \gg T_{-11}.
%\end{equation}
The H1 and ZEUS collaborations have analyzed data in distinct intervals of $Q^2$ and $W$. From here on we will refer only to the H1 ranges, $2.5\,\text{\rm GeV$^2$} < Q^2 <60 \,\text{\rm GeV$^2$}$ and $35\, \text{GeV} < W < 180\,\text{GeV}$
%\begin{equation}
%\begin{split}
%2.5\,\text{\rm GeV$^2$} < Q^2 <60 \,\text{\rm GeV$^2$},\\
%35\, \text{GeV} < W < 180\,\text{GeV},
%\end{split}
%\end{equation}
and will focus only on the dominant helicity amplitude ratio, $T_{11}/T_{00}$. At small $x$ the forward helicity amplitude for the $\rho$-meson leptoproduction can be expressed, in $\kappa$-factorization, as the convolution of the $\gamma^*\rightarrow \rho$ IF,
$\Phi^{\gamma^*(\lambda_\gamma)\rightarrow\rho(\lambda_\rho)}(\kappa^2,Q^2)$,
with the UGD ${\cal F}(x,\kappa^2)$ and reads
\vspace{-0.2cm}
\begin{equation}
\label{amplitude}
T_{\lambda_\rho\lambda_\gamma}(s,Q^2) = \frac{is}{(2\pi)^2}\int \dfrac{d^2\kappa}
{(\kappa^2)^2}\Phi^{\gamma^*(\lambda_\gamma)\rightarrow\rho(\lambda_\rho)}(\kappa^2,Q^2)
{\cal F}(x,\kappa^2),\quad \text{$x=\frac{Q^2}{s}$}\,.
\end{equation}
The definitions of the IFs, for the longitudinal and the transverse cases, assume the form given by Eq.~(33) and Eq.~(38) in~\cite{Anikin:2009bf}. Peculiarly, the longitudinal IF embodies the twist-2 distribution amplitude (DA)~\cite{Ball:1998sk}; while the transverse IF is expressed through DAs which embodies both genuine twist-3 and Wandzura-Wilczek (WW) contributions~\cite{Ball:1998sk, Anikin:2011sa}. %In this work our attention is devoted to the WW contributions and, only in a second step, we release this approximation encompassing the genuine terms. In addition, we employ the \emph{asymptotic} choice for the twist-2 DA (see Sec.~2.2 of~\cite{Bolognino:2018}). 
\vspace{-0.3cm}
\section{Models of unintegrated gluon distribution}
\label{models}
Pursuing the goal to compare distinct approaches, we deal with a collection of six models, introducing here only the functional form ${\cal F}(x,\kappa^2)$ of the UGD. We refer the reader to the original papers for a detailed treatment about the derivation of each model.\\

\textbf{An $x$-independent model (ABIPSW)}\\
An expression for the proton impact factor~\cite{Anikin:2011sa} provides a very simple and $x$-independent UGD: ${\cal F}(x,\kappa^2)= \frac{A}{(2\pi)^2\,M^2}
\left[\frac{\kappa^2}{M^2+\kappa^2}\right]$,
%\begin{equation}
%{\cal F}(x,\kappa^2)= \frac{A}{(2\pi)^2\,M^2}
%\left[\frac{\kappa^2}{M^2+\kappa^2}\right]\,,
%\end{equation}
where $M$ represents the non-perturbative hadronic scale.\\

\textbf{Gluon momentum derivative}\\
This model is given by ${\cal F}(x, \kappa^2) = \frac{dxg(x, \kappa^2)}{d\ln \kappa^2}$
%\begin{equation}
%\label{xgluon}
%{\cal F}(x, \kappa^2) = \frac{dxg(x, \kappa^2)}{d\ln \kappa^2}
%\end{equation}
and encloses the collinear gluon density $g(x, \mu_F^2)$, fixed at
$\mu_F^2=\kappa^2$.\\

\textbf{Ivanov--Nikolaev' (IN) UGD: a soft-hard model}\\
In the large-$\kappa$ range, DGLAP parametrizations for $g(x, \kappa^2)$ are used in this model. Furthermore, for the extrapolation of the hard gluon densities to small $\kappa^2$, an Ansatz is proposed~\cite{Nikolaev:1994cd}. The gluon density at small $\kappa^2$ is endowed with a nonperturbative soft part. This model has the form
\begin{equation}
{\cal F}(x,\kappa^2)= {\cal F}^{(B)}_\text{soft}(x,\kappa^2) 
{\kappa_{s}^2 \over 
	\kappa^2 +\kappa_{s}^2} + {\cal F}_\text{hard}(x,\kappa^2) 
{\kappa^2 \over 
	\kappa^2 +\kappa_{h}^2}\,.
\label{eq:4.7}
\end{equation}
 We refer the reader to~\cite{Ivanov:2000cm} for a meticulous treatment of the parameters and on the expressions for the soft and the hard components.\\
 
\textbf{Hentschinski-Salas--Sabio Vera' (HSS) model}\\
The application of this model occurs, originally, in the study of DIS structure functions~\cite{Hentschinski:2012kr}. Subsequently, it has been used in the description of single-bottom quark production at LHC in~\cite{Chachamis:2015ona}, in the investigation of the photoproduction of $J/\Psi$ and $\Upsilon$ in~\cite{Bautista:2016xnp} and in forward Drell-Yan dilepton production~\cite{Celiberto:2018muu}. This UGD takes the form of a convolution between the BFKL gluon Green's function and a LO proton impact factor:
\begin{equation}
\label{HentsUGD}
{\cal F}(x, \kappa^2, M_h) = \int_{-\infty}^{\infty}
\frac{d\nu}{2\pi^2}\ {\cal C} \  \frac{\Gamma(\delta - i\nu -\frac{1}{2})}
{\Gamma(\delta)}\ \left(\frac{1}{x}\right)^{\chi\left(\frac{1}{2}+i\nu\right)}
\left(\frac{\kappa^2}{Q^2_0}\right)^{\frac{1}{2}+i\nu}
\end{equation}
\[
\times \left\{ 1 +\frac{\bar{\alpha}^2_s \beta_0 \chi_0\left(\frac{1}{2}
	+i\nu\right)}{8 N_c}\log\left(\frac{1}{x}\right)
\left[-\psi\left(\delta-\frac{1}{2} - i\nu\right)
-\log\frac{\kappa^2}{M_h^2}\right]\right\}\,,
\]
where $\chi_0(\frac{1}{2} + i\nu)$ and $\chi(\gamma)$ are respectively the LO and the NLO eigenvalues of the BFKL kernel. We adopted here the so called {\em kinematically improved} values for the parameters $Q_0$, $\delta$ and $\cal{C}$ describing the proton impact factor, (for further details see Sec.~III A of~\cite{Bolognino:2018}).\\

\textbf{Golec-Biernat--W{\"u}sthoff' (GBW) UGD}\\
This type of UGD comes from the effective dipole cross section
$\sigma(x,r)$ for the scattering of a $q\bar{q}$ pair off a
nucleon~\cite{GolecBiernat:1998js}, through a Fourier transform:
\begin{equation}
{\cal F}(x,\kappa^2)= \kappa^4 \sigma_0 \frac{R^2_0(x)}{8\pi}
e^{\frac{-k^2 R^2_0(x)}{4}}\,.
\end{equation}
We refer to~\cite{GolecBiernat:1998js} for insights and discussion of the parameters of this model.\\

\textbf{Watt--Martin--Ryskin' (WMR) model}\\
The UGD proposed in~\cite{Watt:2003mx} reads
\vspace{-0.4cm}
\[
{\cal F}(x,\kappa^2,\mu^2) = T_g(\kappa^2,\mu^2)\,\frac{\alpha_s(\kappa^2)}
{2\pi}\,\int_x^1\!dz\;\left[\sum_q P_{gq}(z)\,\frac{x}{z}q\left(\frac{x}{z},
\kappa^2\right) + \right.\nonumber
\]
\vspace{-0.25cm}
\begin{equation}
\label{WMR_UGD}
\left. \hspace{5.3cm} P_{gg}(z)\,\frac{x}{z}g\left(\frac{x}{z},\kappa^2\right)\,\Theta\left(\frac{\mu}{\mu+\kappa}-z\right)\,\right]\,,
\end{equation}
where the term $T_g(\kappa^2,\mu^2)$, whose expression is provided in~\cite{Watt:2003mx}, indicates the probability of evolving from the scale $\kappa$ to the scale $\mu$ without parton emission.
This UGD model depends on an extra-scale $\mu$, fixed at $Q$.
\begin{figure}[h]
	\centering
	\includegraphics[scale=0.264,clip]{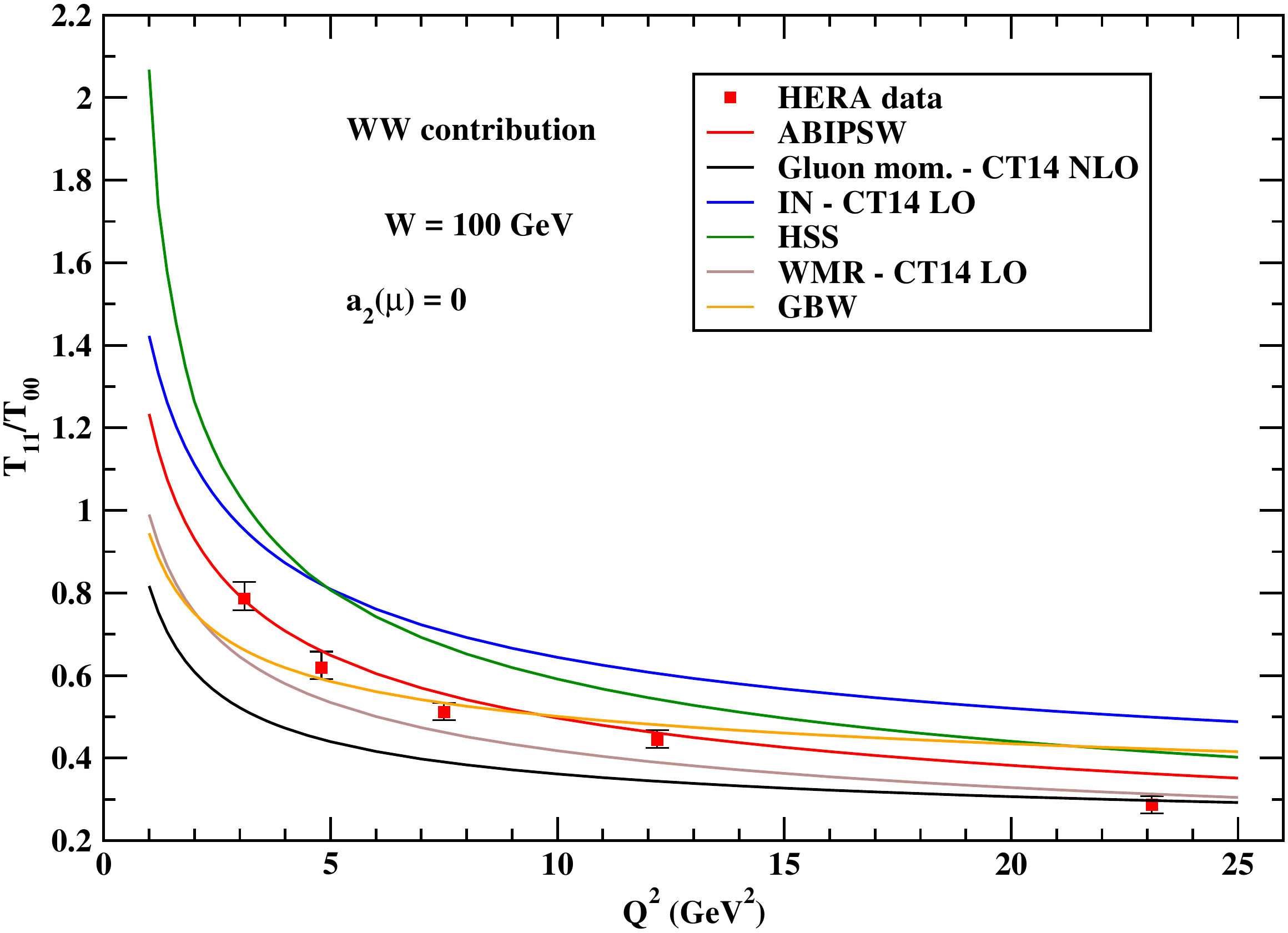}
	\caption{\emph{$Q^2$-dependence of $T_{11}/T_{00}$ for
			the six UGD models at $W = 100$ GeV.}}
	\label{fig:ratio_all}
\end{figure}
\begin{figure}[h]
	\centering
	\includegraphics[scale=0.264,clip]{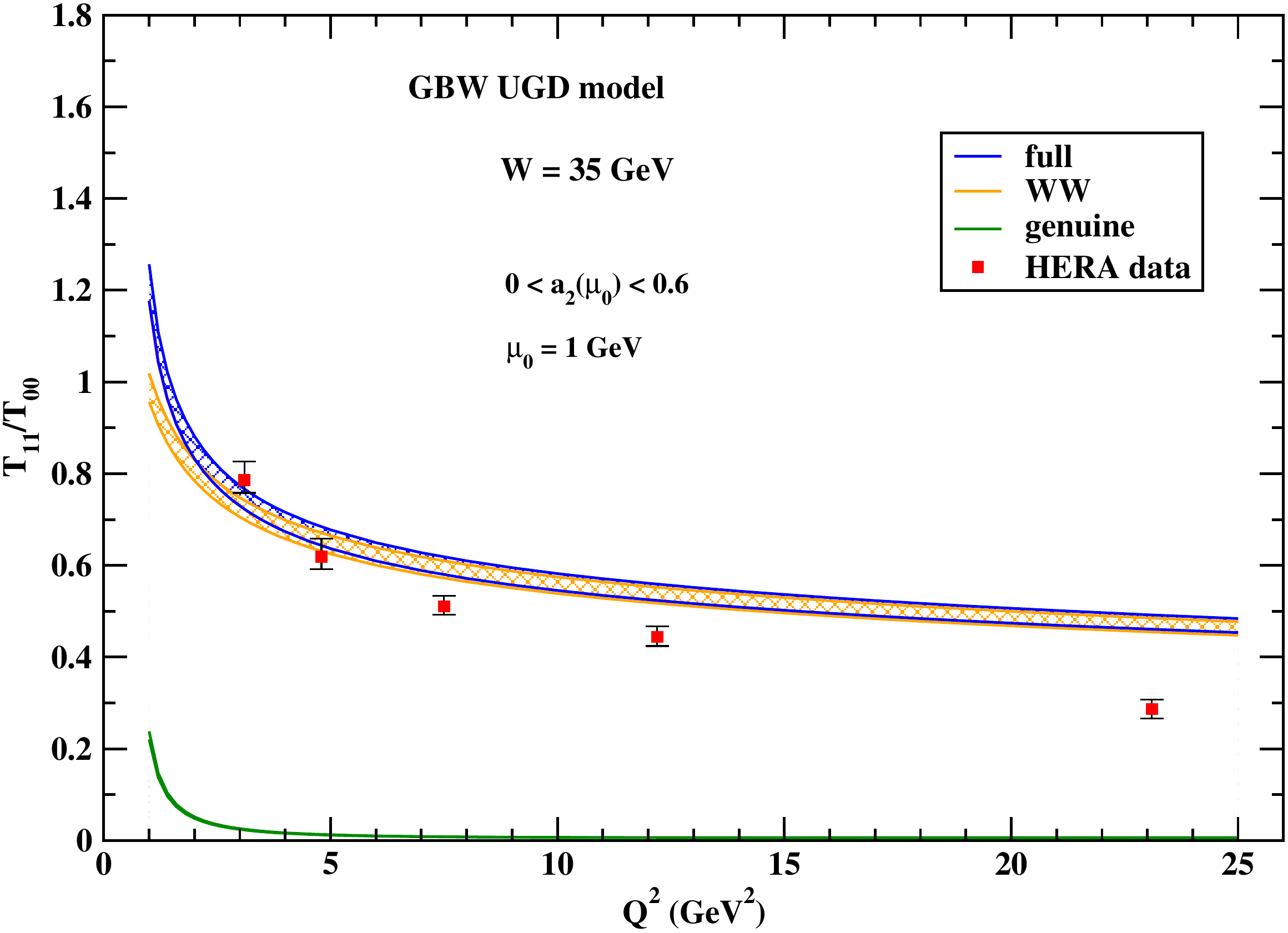}	
	\includegraphics[scale=0.264,clip]{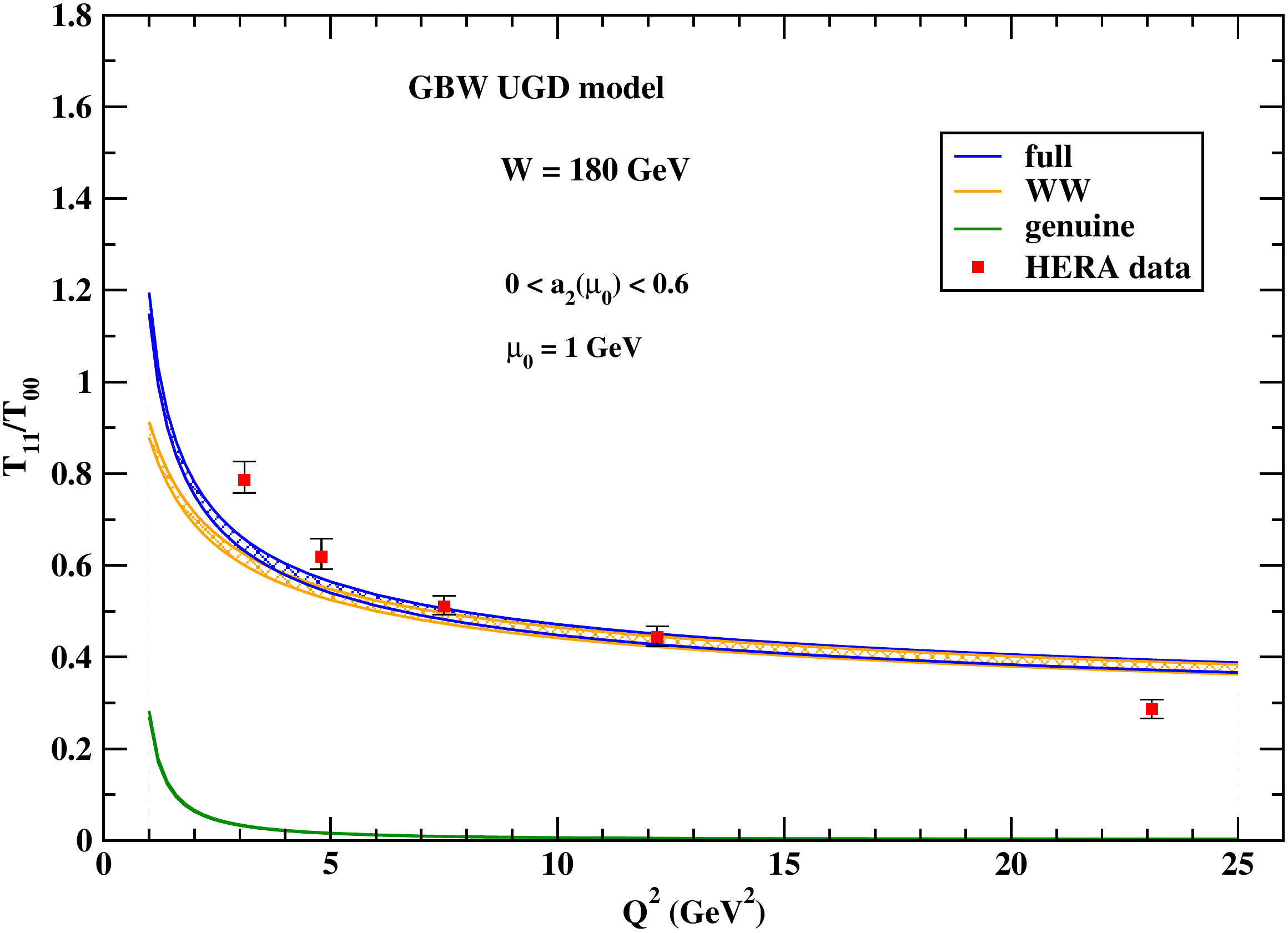}
	\caption{\emph{$Q^2$-dependence of $T_{11}/T_{00}$ for
			the GBW UGD model at $W = 35$ (left) and 180~GeV (right). The full, WW and
			genuine contributions are shown. The bands give the effect of
			varying $a_2(\mu_0 = 1 \mbox{ GeV})$ between 0. and 0.6.}}
	\label{fig:ratio_GBW_evolved}
\end{figure}
\begin{figure}[h]
	\centering	
	\includegraphics[scale=0.264,clip]{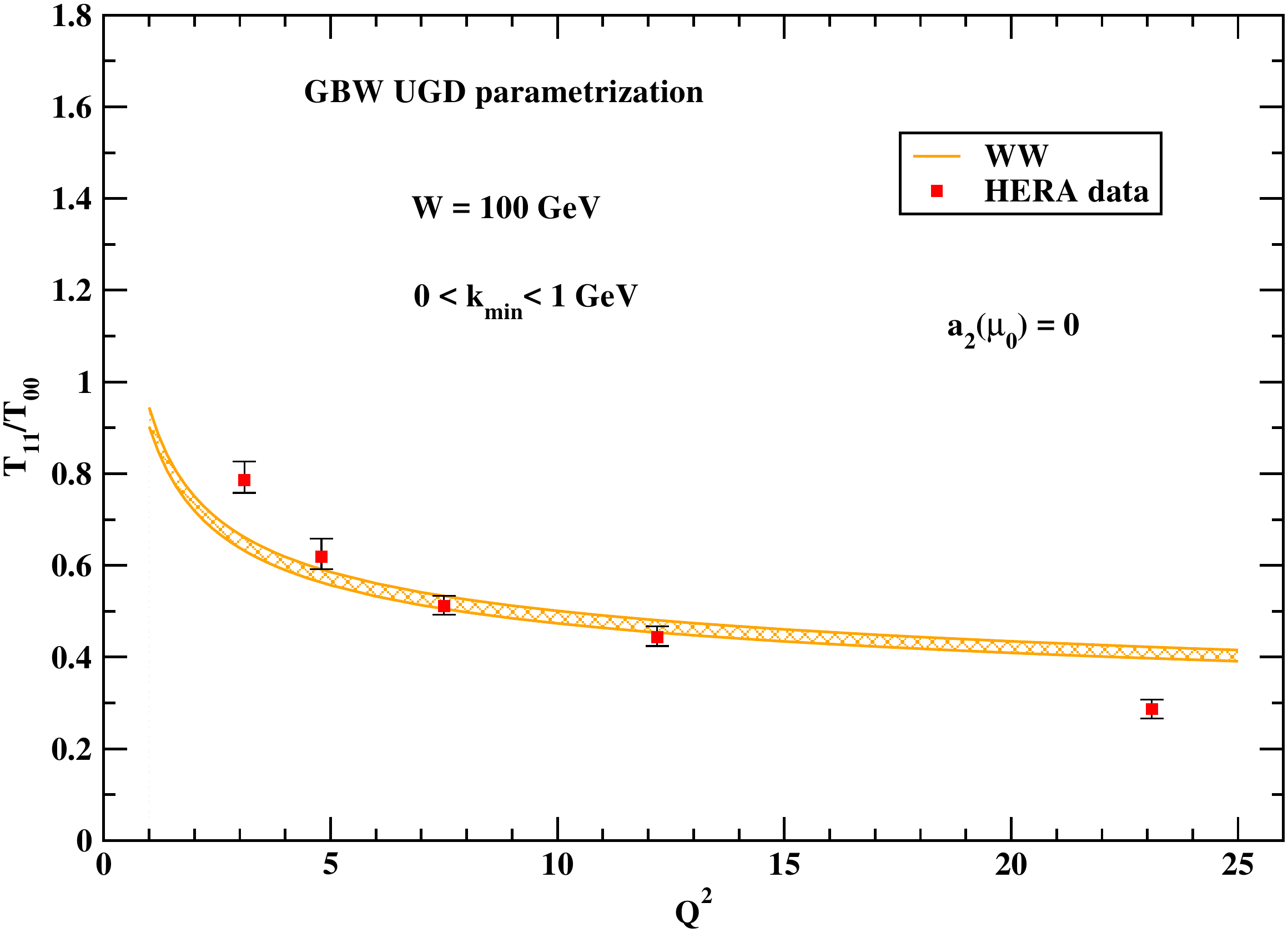}
	\caption{\emph{$Q^2$-dependence of $T_{11}/T_{00}$ for
			the GBW UGD model at $W = 100$~GeV. The band is the effect of a lower
			cutoff in the $\kappa$-integration, ranging from 0. to 1~GeV.}}
	\label{fig:ratio_GBW_kmin}
\end{figure}
%\vspace{-0.6cm}
\section{Numerical analysis}
%\vspace{-0.05cm}
We propose our predictions for the helicity-amplitude ratio $T_{11}/T_{00}$, as obtained with the selection of six UGD models presented in Sec.~\ref{models}, and compare them with HERA data. Fig.~\ref{fig:ratio_all} exhibits the comparison between the $Q^2$-dependence of $T_{11}/T_{00}$ for all six models at $W = 100$~GeV and the experimental result. We exploited here the \emph{asymptotic} twist-2 DA ($a_2(\mu^2)=0$) and the WW approximation for twist-3 contributions. The fact that the $T_{11}/T_{00}$ is measured on a large $Q^2$-interval allows to strongly constrain the $\kappa$ dependence of UGDs. None of the models is able to describe data over the entire $Q^2$ range; only the $x$-independent ABIPSW model and the GBW model seem to better catch the intermediate-$Q^2$ behavior of data. In order to calibrate the effect of the approximation made in the DAs, we performed the $T_{11}/T_{00}$ ratio with the GBW model, at $W = 35$ and 180~GeV,
by varying $a_2(\mu_0=1 \ {\rm GeV})$ in the range 0. to 0.6 and suitably taking into account its evolution. Besides, for the same UGD model as we observe in Fig.~\ref{fig:ratio_GBW_evolved}, we relaxed the WW approximation in $T_{11}$ and examined also the genuine twist-3 contribution. This plot illustrates that our predictions for $T_{11}/T_{00}$ are rather insensitive to the form of the meson DAs. The stability of $T_{11}/T_{00}$ under the lower cut-off for $\kappa$, in the range 0 $< \kappa_{\rm min} < 1$~GeV, has been probed. The result of this test is illustrated in Fig.~\ref{fig:ratio_GBW_kmin} for the GBW model at $W = 100$~GeV. It is clear that the small-$\kappa$ region gives only a marginal contribution. This is a crucial point because it supports the main underlying assumption of this work, namely that {\em both} the helicity amplitudes, $T_{11}$ and $T_{00}$, are dominated by the large-$\kappa$ region.
%\vspace{-0.5cm}
\section{Conclusions}
\label{discussion}
%\vspace{-0.05cm}
We have proposed the helicity amplitudes for the leptoproduction of vector mesons at HERA as a nontrivial testing ground for models of the UGD in the proton. We have provided some theoretical arguments that both the transverse and the longitudinal case are dominated by the kinematic region where small-size color dipoles interact with the proton. Furthermore, this investigation shows that some of the most popular models for the UGD in the literature give very sparse predictions for the ratio $T_{11}/T_{00}$.

%uncomment the following lines to place a figure
%\begin{figure}[htb]
%\centerline{%
%\includegraphics[width=12.5cm]{Fig1}}
%\caption{Plot of ...}
%\label{Fig:F2H}
%\end{figure}
%\vspace{-0.7cm}


\begin{thebibliography}{99}
%\vspace{-0.4cm}
	\bibitem{BFKL}
	V.S.~Fadin {\it et al.}, Phys. Lett. {\bf B60} (1975) 50;
	E.A.~Kuraev {\it et al.}, Zh. Eksp. Teor. Fiz. {\bf 71} (1976)
	840 [Sov. Phys. JETP {\bf 44} (1976) 443]; {\bf 72} (1977) 377 [{\bf 45} (1977)
	199];
	Ya.Ya.~Balitskii, L.N.~Lipatov, Sov. J. Nucl. Phys. {\bf 28} (1978) 822.
	
	\bibitem{small_x_WG}
	J.R.~Andersen {\it et al.} [Small x Coll.],
	Eur. Phys. J. C {\bf 48} (2006) 53; %[hep-ph/0604189];
	Eur. Phys. J. C {\bf 35} (2004) 67; %[hep-ph/0312333];
	B.~Andersson {\it et al.} [Small x Coll.],
	Eur. Phys. J. C {\bf 25} (2002) 77.% [hep-ph/0204115].

	%\cite{Bolognino:2018}
	\bibitem{Bolognino:2018}
	A.D. Bolognino, F.G. Celiberto, D.Yu.~Ivanov, A. Papa, Eur.\ Phys.\ J.\ C {\bf 78} (2018) no.12, 1023.
	
	%\cite{Bolognino:2018mlw}
	\bibitem{Bolognino:2018mlw}
	A.D.~Bolognino, F.G.~Celiberto, D.Yu.~Ivanov, A.~Papa,
	%``$\rho$-meson leptoproduction as testfield for the unintegrated gluon distribution in the proton,''
	[arXiv:1808.02958 [hep-ph]].
	%%CITATION = ARXIV:1808.02958;%%
	
	%\cite{Aaron:2009xp}
	\bibitem{Aaron:2009xp}
	F.D.~Aaron {\it et al.} [H1 Coll.],
	%``Diffractive Electroproduction of rho and phi Mesons at HERA,''
	JHEP {\bf 1005} (2010) 032.
	% doi:10.1007/JHEP05(2010)032
	%[arXiv:0910.5831 [hep-ex]].
	
	%\cite{Ivanov:1998gk}
	\bibitem{Ivanov:1998gk}
	D.Yu.~Ivanov, R.~Kirschner,
	%``Polarization in diffractive electroproduction of light vector mesons,''
	Phys.\ Rev.\ D {\bf 58} (1998) 114026.
	%[hep-ph/9807324].
	
	%\cite{Ball:1998sk}
	\bibitem{Ball:1998sk}
	P.~Ball {\it et al.},
	%``Higher twist distribution amplitudes of vector mesons in QCD: Formalism and twist - three distributions,''
	Nucl.\ Phys.\ B {\bf 529} (1998) 323.
	% doi:10.1016/S0550-3213(98)00356-3
	%[hep-ph/9802299].
	
	%\cite{Anikin:2009bf}
	\bibitem{Anikin:2009bf}
	I.V.~Anikin {\it et al.},
	%``QCD factorization of exclusive processes beyond leading twist: gamma*T ---> rho(T) impact factor with twist three accuracy,''
	Nucl.\ Phys.\ B {\bf 828} (2010) 1.
	% doi:10.1016/j.nuclphysb.2009.10.022
	%[arXiv:0909.4090 [hep-ph]].
	%	\vspace{-0.05cm}
	%\cite{Anikin:2011sa}
	\bibitem{Anikin:2011sa}
	I.V.~Anikin {\it et al.},
	%``A phenomenological study of helicity amplitudes of high energy exclusive leptoproduction of the rho meson,''
	Phys.\ Rev.\ D {\bf 84} (2011) 054004.
	%  doi:10.1103/PhysRevD.84.054004
	%[arXiv:1105.1761 [hep-ph]].
	%%CITATION = doi:10.1103/PhysRevD.84.054004;%%
	%20 citations counted in INSPIRE as of 09 Mar 2018
	
	
	%\cite{Nikolaev:1994cd}
	\bibitem{Nikolaev:1994cd}
	N.N.~Nikolaev, B.G.~Zakharov,
	%``Splitting the pomeron into two jets: A Novel process at HERA,''
	Phys.\ Lett.\ B {\bf 332} (1994) 177;
	%  doi:10.1016/0370-2693(94)90876-1
	%[hep-ph/9403281];
	Z.\ Phys.\ C {\bf 53} (1992) 331.
	%%CITATION = doi:10.1016/0370-2693(94)90876-1;%%
	%74 citations counted in INSPIRE as of 09 Mar 2018
	
	%\cite{Ivanov:2000cm}
	\bibitem{Ivanov:2000cm}
	I.P.~Ivanov, N.N.~Nikolaev,
	%``Anatomy of the differential gluon structure function of the proton from the experimental data on F(2p) (x, Q**2),''
	Phys.\ Rev.\ D {\bf 65} (2002) 054004.
	%  doi:10.1103/PhysRevD.65.054004
	%[hep-ph/0004206].
	%%CITATION = doi:10.1103/PhysRevD.65.054004;%%
	%90 citations counted in INSPIRE as of 09 Mar 2018
	

	%\cite{Hentschinski:2012kr}
	\bibitem{Hentschinski:2012kr}
	M.~Hentschinski {\it et al.},
	%``Hard to Soft Pomeron Transition in Small-x Deep Inelastic Scattering Data Using Optimal Renormalization,''
	Phys.\ Rev.\ Lett.\  {\bf 110} (2013) no.4, 041601.
	%  doi:10.1103/PhysRevLett.110.041601
	%[arXiv:1209.1353 [hep-ph]].
	%%CITATION = doi:10.1103/PhysRevLett.110.041601;%%
	%58 citations counted in INSPIRE as of 09 Mar 2018
	
	%\cite{Chachamis:2015ona}
	\bibitem{Chachamis:2015ona}
	G.~Chachamis {\it et al.},
	%``Single bottom quark production in k$_{?}$-factorisation,''
	JHEP {\bf 1509} (2015) 123.
	%  doi:10.1007/JHEP09(2015)123
	%[arXiv:1507.05778 [hep-ph]].
	%%CITATION = doi:10.1007/JHEP09(2015)123;%%
	%12 citations counted in INSPIRE as of 09 Mar 2018
	
	%\cite{Bautista:2016xnp}
	\bibitem{Bautista:2016xnp}
	I.~Bautista {\it et al.},
	%``BFKL evolution and the growth with energy of exclusive $J/\psi$ and $\Upsilon$ photoproduction cross sections,''
	Phys.\ Rev.\ D {\bf 94} (2016) no.5,  054002.
	%  doi:10.1103/PhysRevD.94.054002
	%[arXiv:1607.05203 [hep-ph]].
	%%CITATION = doi:10.1103/PhysRevD.94.054002;%%
	%8 citations counted in INSPIRE as of 09 Mar 2018
	
	\bibitem{Celiberto:2018muu}
	F.G.~Celiberto {\it et al.},
	%``Forward Drell–Yan production at the LHC in the BFKL formalism with collinear corrections,''
	Phys.\ Lett.\ B {\bf 786} (2018) 201.
	%[arXiv:1808.09511 [hep-ph]].
	
	%\cite{GolecBiernat:1998js}
	\bibitem{GolecBiernat:1998js}
	K.J.~Golec-Biernat, M.~W\"usthoff,
	%``Saturation effects in deep inelastic scattering at low Q**2 and its implications on diffraction,''
	Phys.\ Rev.\ D {\bf 59} (1998) 014017.
	%  doi:10.1103/PhysRevD.59.014017
	%
	%[hep-ph/9807513].
	%%CITATION = doi:10.1103/PhysRevD.59.014017;%%
	%1127 citations counted in INSPIRE as of 09 Mar 2018
		%\vspace{-0.05cm}
	%\cite{Watt:2003mx}
	\bibitem{Watt:2003mx}
	G.~Watt {\it et al.},
	%``Unintegrated parton distributions and inclusive jet production at HERA,''
	Eur.\ Phys.\ J.\ C {\bf 31} (2003) 73.
	%  doi:10.1140/epjc/s2003-01320-4
	%[hep-ph/0306169].
	%%CITATION = doi:10.1140/epjc/s2003-01320-4;%%
	%139 citations counted in INSPIRE as of 09 Mar 2018
	
	%	\bibitem{cernlib}
	%	%CERNLIB Homepage: \url{http://cernlib.web.cern.ch/cernlib}.
	
	%	\bibitem{Harland-Lang:2014zoa}
	%L.A.~Harland-Lang, A.D.~Martin, P.~Motylinski, R.S.~Thorne,
	%``Parton distributions in the LHC era: MMHT 2014 PDFs,''
	%	Eur.\ Phys.\ J.\ C {\bf 75} (2015) no.5,  204
	%  doi:10.1140/epjc/s10052-015-3397-6
	%[arXiv:1412.3989 [hep-ph]].
	
	%\cite{Ball:2014uwa}
	%	\bibitem{Ball:2014uwa}
	%	R.D.~Ball {\it et al.} [NNPDF Collaboration],
	%``Parton distributions for the LHC Run II,''
	%	%JHEP {\bf 1504} (2015) 040
	%  doi:10.1007/JHEP04(2015)040
	%	[arXiv:1410.8849 [hep-ph]].
	%%CITATION = doi:10.1007/JHEP04(2015)040;%%
	%1036 citations counted in INSPIRE as of 09 Mar 2018
	
	%\cite{Buckley:2014ana}
	%\bibitem{Buckley:2014ana}
	%A.~Buckley, J.~Ferrando, S.~Lloyd, K.~Nordstr{\"o}m, B.~Page, M.~R{\"u}fenacht, M.~Schönherr, G.~Watt,
	%``LHAPDF6: parton density access in the LHC precision era,''
	%Eur.\ Phys.\ J.\ C {\bf 75} (2015) 132
	%  doi:10.1140/epjc/s10052-015-3318-8
	%[arXiv:1412.7420 [hep-ph]].
	%%CITATION = doi:10.1140/epjc/s10052-015-3318-8;%%
	%307 citations counted in INSPIRE as of 09 Mar 2018
%	\vspace{-0.2cm}
\end{thebibliography}
\end{document}